\documentstyle[12pt,epsf]{article}
\begin{document}

\newcommand{\qq}{ Q^2}
\newcommand{\alst}{ \alpha_{\rm s}}
\newcommand{\rd}{{\rm d}}
\newcommand{\rln}{{\rm ln}}
\newcommand{\rLi}{{\rm Li}}
\newcommand{\GeV}{{\rm GeV}}
\newcommand{\NS}{{\rm NS}}
\newcommand{\Si}{{\rm S}}
\newcommand{\gw}{ g_1}
\newcommand{\Dqns}{\Delta q^{\rm NS}}
\newcommand{\Dsgm}{\Delta\Sigma}

%%%%%%%%%%%%%%%%%%%%%%%%%%%%%%%%%%%%%%%%%%%%%%%%%%%%%%%%%%%%%%%%%%%%%%%%%%%%

\begin{titlepage}
\begin{flushright}
\today
\end{flushright}
\begin{center}
\vskip 1cm
{\Large\bf A (Slightly Less Brutal) Method for Numerically Evaluating
        Structure Functions}
\vskip 2cm
{\large D.~Fasching}
\vskip .8cm
{\it Physics Department, University of Wisconsin-Madison, 1150 University
     Avenue, Madison, Wisconsin 53706, USA}$^\dagger$
\vskip 1cm
\end{center}
 
\begin{abstract}
A fast numerical algorithm for the evolution of parton distributions in $x$
space is described.  The method is close in spirit to `brute' force
techniques.  The necessary integrals are performed by summing the approximate
contributions
from small steps of the integration region.
Because it is a numerical evaluation it shares the advantage with
brute force numerical integration that
there are no restrictions placed on the functional form of the distributions
to be evolved.  However, an improvement in the approximation technique
results in a significant
reduction in the number of integration steps and a savings in time
on the order of three hundred fifty.
The method has been implemented for the structure functions $F_2$ and $g_1$ at
next-to-leading order.
\end{abstract}

\vskip 4cm
\hskip -6mm\line(1,0){170}
\newline
\hskip -6mm$^\dagger$Currently located at Lawrence Berkeley National Laboratory,
MS 50-348, 1 Cyclotron Road, Berkeley CA 94720, USA.

\vfill
\end{titlepage}

\newpage

\section{Introduction}

Much of the information which can be gained from studying collisions at
high energy hadron colliders can only be extracted from the data if one has
previous knowledge of hadronic structure.  Presently this knowledge
is not available by direct calculation and so must be obtained by other means.
Dedicated high energy lepton-proton deep inelastic scattering, DIS, experiments
provide the means to measure this structure.  The structure functions which
are measured in such experiments are essentially the cross-sections and
hence are not directly applicable to other processes.  However,
they can be related to parton
distributions which are universal and which can therefore be used to predict,
for example, production rates in the more complicated environment of a hadron
collider.  There are other reasons why determining parton distributions is of
interest.  For example, one hopes that at some point they will
be accessible by direct calculation.  Such predictions will then need to be
compared with
experimental measurements.  In addition, the proper evaluation of DIS
sum rules relies on the same formalism by which the parton
distributions are extracted.  These reasons, along with the large and growing
body of DIS data, make a strong case for efficient tools for determining
parton distributions.

Structure functions are related to parton distributions via the operator
product expansion.  Within this framework, the structure function at some
scale, $\qq$, is expressed in terms of a convolution of parton distributions
with coefficient functions which can be calculated perturbatively
\cite{wilson_69}.  For the case of charged lepton scattering in the one
photon exchange approximation we have,
\begin{equation}
 F_i(x,\qq)=a_i(x)\left\{<e^2>\left[\Sigma(\qq)\otimes_xC_i^\Sigma +
                           G(\qq)\otimes_xC_i^G\right] +
                         q^{NS}(\qq)\otimes_xC_i^{(NS)}\right\},
 \label{eq:str_fcn}
\end{equation}
where $f\otimes_xq\equiv\int_x^1\frac{\rm dz}{z}f\left(\frac{x}{z}\right)q(z)$,
$a_2 = a_L = x$, $a_1 = \frac{1}{2}$,
$\Sigma(x,\qq)$, $G(x,\qq)$ and $q^{NS}(x,\qq)$ are the
quark singlet, gluon and nonsinglet distributions, respectively, the $C_i$ are
the coefficient functions and $<e^2>$
is the mean squared charge of the involved quarks.
In (\ref{eq:str_fcn}) the factorization scale and the scale at which the
structure function is defined have been implicitly set equal to each other.
Because this will always be the case in this paper, the more general expression
is not given.  For $g_1$, $a_1 = \frac{1}{2}$ and the parton distributions are
substituted by their polarized counterparts, $\Delta\Sigma$, $\Delta G$ and
$\Delta q^{NS}$.  Eq. (\ref{eq:str_fcn}) can be taken as a definition
of the parton distributions.  The distributions, therefore, depend on the
details of the calculation of the coefficient functions, which are not unique.

Though the parton distributions can not presently be calculated, their
dependence on
$\qq$ is predictable for high enough values of $\qq$.  Because of this, the
large body of DIS data
currently available, extending up to $5\times 10^4$ GeV$^2$ and down
to $x = 10^{-5}$, can
be used together in a single global fit of the distributions.  Their $\qq$
dependence is given by the DGLAP equations \cite{dglap}.  Because the
QCD processes which govern this dependence do not act on flavor indices,
the (flavor) singlet distribution, in which these indices have been summed
out, and the nonsinglet distribution behave differently.
In particular, the gluon distribution, which also does not carry flavor,
can have no effect on the nonsinglet distribution in the perturbative evolution.
Thus, in the nonsinglet sector we have
\begin{equation}
 \frac{\rm{d}}{\rm{d}t}q^{NS}(x,t) = \frac{\alpha_s(t)}{4\pi}
                   P_{qq}^{NS}\biggl(\alpha_s(t)\biggr)\otimes_x q^{NS}(t),
 \label{eq:nonsinglet}
\end{equation}
where $t = \rm{ln}\frac{\qq}{\Lambda^2}$ and $P^{NS}(x,\alpha_s(\qq))$ is the
nonsinglet quark splitting function,
which can be calculated in perturbative QCD.
The coupled evolution equations in the singlet sector are
\begin{eqnarray}
 \frac{\rm{d}}{\rm{d}t}\Sigma(x,t) & = & \frac{\alpha_s(t)}{4\pi}
        \left[P_{qq}^\Sigma\biggl(\alpha_s(t)\biggr)\otimes_x \Sigma(t) +
        P_{qg}\biggl(\alpha_s(t)\biggr)\otimes_x G(t)\right], \nonumber \\
 \frac{\rm{d}}{\rm{d}t}G(x,t) & = & \frac{\alpha_s(t)}{4\pi}
        \left[P_{gq}\biggl(\alpha_s(t)\biggr)\otimes_x \Sigma(t) +
        P_{gg}\biggl(\alpha_s(t)\biggr)\otimes_x G(t)\right].
 \label{eq:singlet}
\end{eqnarray}
The present state of knowledge of the splitting functions and coefficient
functions allows for a next-to-leading order, NLO, treatment of both
the polarized and
unpolarized structure functions.  For a concise review of the state of the
art of these calculations see reference \cite{neerven_review}.

The relationship (\ref{eq:str_fcn}-\ref{eq:singlet}) between
the objects which are parameterized, the parton distributions at a
fixed scale, and the data to which they are fit, structure function data
over a wide kinematic range, is quite complicated.  In fact, these equations
do not have analytic solutions.  Because of this several techniques, including
integral transforms, polynomial expansions and numerical methods, have been
employed to obtain approximate solutions.

The most popular of the polynomial expansion techniques involves expanding
the splitting and coefficient functions, as well as the parton distributions,
in Laguerre polynomials \cite{furmanski_81,ramsey_84}.
In this technique, the convolutions
(\ref{eq:str_fcn}-\ref{eq:singlet}) reduce to
a sum of products of Laguerre polynomials.  A study of the Laguerre
method \cite{kumano_91} showed that one could obtain accurate results quickly
in the range $0.01 < x < 0.8$.  Outside of this range, it was observed that
the Laguerre expansion no longer describes the functions and so the method
breaks down.  Given that polarized scattering data is available down
to $x = 3\times 10^{-3}$ \cite{smc} and unpolarized data is available
down to $x=10^{-5}$ \cite{h1,zeus}, this method is no longer practical.

Another standard procedure is to transform the entire set of equations into
moment space via a Mellin transformation.
The convolutions of the $x$ dependent functions appear as products of their
moment space counterparts.  This simpler set of equations has a closed form
solution which can be transformed back to $x$ space for comparison with data.
This technique is typically much faster than a direct numerical evaluation in
$x$ space, which is another popular technique.

A `brute force' numerical integration method was discussed in references
\cite{ramsey_phd,kumano_95}.
In this method the convolutions in Eqs. (\ref{eq:str_fcn}-\ref{eq:singlet})
are performed by straightforward numerical integration, dividing the
integration region into small steps and
summing their approximate contributions to the integral,
\begin{eqnarray}
  P\otimes_xq & = &
         \sum_{j}\int_{x_j}^{x_{j+1}}
                \frac{{\rm d}z}{z}P(z)q\left(\frac{x}{z}\right)_j,
\label{eq:steps} \\
 & \approx & \sum_{j}
     \frac{\delta x_j}{x_{j+1/2}}P(x_{j+1/2})q\left(\frac{x}{x_{j+1/2}}\right).
\label{eq:brutal}
\end{eqnarray}
Here, $P$ represents a general splitting or coefficient function, $q$ is a general
parton distribution,
\mbox{$\delta x_j = x_{j+1} - x_j$ and $x_{j+1/2}$} lies somewhere between
$x_j$ and $x_{j+1}$.  Similarly, the differential in $t$ is approximated by a
finite difference.  Eq. (\ref{eq:nonsinglet}) becomes
\begin{equation}
 \Dqns(x,t_{l\pm1})\approx\Dqns(x,t_l) + \delta t_l\cdot\frac{\alst(t_l)}{4\pi}
    \cdot\sum_{j}\frac{\delta x_j}{x_{j+1/2}}
  P\left(x_{j+1/2},\alst(t_l)\right)\cdot q\left(\frac{x}{x_{j+1/2}},t_l\right),
\end{equation}
where $\delta t_l = t_{l\pm1}-t_l$.  Similar expressions hold in the singlet
sector.  No restrictions are
imposed on the functional form of the distributions because all of the functions
are evaluated numerically.
For the same reason, as was pointed out and
in fact implemented in reference \cite{kumano_95},
it is well suited for solving the nonlinear modified
evolution equations \cite{mueller_87} which take recombination effects into
account.

Because the integral on the left hand side of Eq.
(\ref{eq:steps}) must be performed with each $x$ step value as the
lower limit, the number of times the integral on the right hand side must
be performed is roughly proportional to the square of the number of $x$ steps.
This must be repeated for each point in the $t$ grid.
For the implementation of reference \cite{kumano_95}
an accuracy of 2\% on unpolarized distributions in the kinematic range
relevant to the HERA collider experiments requires 1000 steps in ln($x$),
with a running time of roughly one hour.\footnote[1]{The program was run on
an AlphaServer 2100 4/200.}
This is acceptable to evolve a fixed set of
distributions but not practical for performing fits of parton distributions
where the evolution may need to be repeated several hundred times.
In addition, the freedom to try different functional forms, to try different
data sets, or to perform other checks on the fit result is not easily
accommodated by such a time consuming procedure.
Below, a slightly less `brutal' numerical technique is described which has
the advantages of the brute force method but for which the running time is
considerably reduced, making it more useful for performing fits.

\section{Semianalytic Solution}

Up to NLO, the splitting and coefficient functions found in the references
of \cite{neerven_review} can be cast in the form
\begin{equation}
 P(x) = P_a(x) + \sum_{k=0}^{1}P_kf_{k+}(x) + P_\delta\delta(1-x),
 \label{eq:qcd_func}
\end{equation}
where $P_0$, $P_1$ and $P_\delta$ are numerical coefficients and $P_a(x)$ is
a function of $x$.  The `plus' distributions,
\begin{equation}
 f_{k+}(x) = \left(\frac{{\rm ln}^k(1-x)}{1-x}\right)_+,
\end{equation}
are defined via the following integral over the interval 0 to 1,
\begin{equation}
 \int_0^1\left(\frac{\rln^k(1-z)}{1-z}\right)_+ g(z)\rd z =
  \int_0^1\frac{\rln^k(1-z)}{1-z}\left[g(z)-g(1)\right]\rd z.
\label{eq:plus}
\end{equation}
Though the method to be described does not depend on the precise form of the
splitting and coefficient functions, it is instructive to separate the
convolution integrals into the pieces suggested by (\ref{eq:qcd_func}).

\subsection{Basic Method and Convolutions with $P_a(x)$}

As in the brute force method, the starting point is Eq. (\ref{eq:steps}).
We would like to approximate the integrand on the right hand side of that
equation as well as possible.  In the brute force method the value of the
integrand throughout the bin is approximated by its value at the bin center.
In what will be called the semianalytic method it is first noted that
approximating the integrals of the splitting and coefficient functions is not
necessary because their functional dependencies are known.
They could, for example, be explicitly
integrated throughout the bin and the result multiplied by the value of the
parton distribution at the bin center.  There is a technical difficulty
associated with this which will be described in the next section.  For
now we simply state that one solution to this difficulty is to instead
approximate the distribution with a linear interpolation in $\frac{1}{z}$
between its values at each pair of adjacent points in the $x$ grid, $z=x_j$ and
$z=x_{j+1}$,
\begin{equation}
 q\left(\frac{x}{z}\right)_j \approx
    \frac{q\left(\frac{x}{x_{j+1}}\right) - q\left(\frac{x}{x_j}\right)}
         {\left(\frac{x}{x_{j+1}} - \frac{x}{x_j}\right)} \cdot
    \left(\frac{x}{z} - \frac{x}{x_j}\right)
 + q\left(\frac{x}{x_j}\right).
\label{eq:interp}
\end{equation}
Substituting this expression into the right hand side of Eq. (\ref{eq:steps})
and taking only the $P_a$ piece of Eq. (\ref{eq:qcd_func}), we arrive at the
expression for the convolution of $P_a(z)$ with $q\left(\frac{x}{z}\right)$ in
the semianalytic method,

\begin{equation}
 P_a\otimes_xq \approx
  \sum_j\biggl\{W_1 \int_{x_j}^{x_{j+1}}\rd z\frac{P_a(z)}{z}
        + W_2 \int_{x_j}^{x_{j+1}}\rd z\frac{P_a(z)}{z^2}\biggr\},
 \label{eq:p0_int}
\end{equation}
where
\begin{eqnarray}
  W_1 & = & \frac{x_{j+1}q\left(\frac{x}{x_{j+1}}\right) 
                              - x_jq\left(\frac{x}{x_j}\right)}
                                    {x_{j+1} - x_j},
 \label{eq:w1}   \\
 & & \nonumber \\
  W_2 & = & -\frac{x_jx_{j+1}}{x_{j+1}-x_j}
    \left[q\left(\frac{x}{x_{j+1}}\right)-q\left(\frac{x}{x_j}\right)\right].
 \label{eq:w2}
\end{eqnarray}
The integrals appearing in (\ref{eq:p0_int}) need only be evaluated once for
each $x$ step.
The $W_k$ must be evaluated for each pairwise combination of $x$ steps
at each $t$ step of the evolution.  If a fit is being performed, the $W_k$
must be reevaluated each time a parameter is changed which affects the
value of $q$.

This method, in addition to retaining all of the information of the splitting
and coefficient functions, improves the approximation of the parton
distributions (which can only be known approximately in any case) compared to
that used in the brute force method.  The resolution of the $x$ grid needed to
obtain a reasonable approximation is determined by the quality of the data.
Beyond a certain resolution, the accuracy of the approximation will exceed what
the data can tell us about the distributions.  In the implementation
of this method described in
\cite{thesis} which treated only $g_1$, the indefinite integrals
$\int{\rm d}z\frac{P_a(z)}{z}$ and $\int{\rm d}z\frac{P_a(z)}{z^2}$ were
explicitly worked out and were also evaluated numerically by gaussian
quadrature.
Differences between the results were negligible, and so in this implementation,
which includes $F_2$ as well, all integrals are performed numerically.
The dilogarithm, $\rm Li_2(x)$, which occurs in the splitting and coefficient
function has been evaluated using its power series form.  Because performing
numerical integrals involving this power series can be
time consuming for $|x|\sim 1$ the substitutions
\begin{eqnarray}
{\rm Li}_2(x)&=&\frac{\pi^2}{6}-{\rm ln}(x){\rm ln}(1-x)-{\rm Li}_2(1-x),
        \nonumber \\
{\rm Li}_2(x)&=&-{\rm Li}_2\left(\frac{-x}{1-x}\right) -
		\frac{1}{2}{\rm ln}^2\left(\frac{1}{1-x}\right), \nonumber
\end{eqnarray}
have been used for $x\sim 1$ and $x\sim -1$, respectively.

\subsection{Convolutions with the Plus Terms}

Apart from a minor technical detail, the plus terms are treated in the same
way.  Starting with the definition in Eq. (\ref{eq:plus}), we can write
\begin{equation}
 f_{k+}\otimes_xq =
  \int_x^1\rd z\frac{\rln^k(1-z)}{1-z}\left[\frac{q\left(\frac{x}{z}\right)}{z}
                        - q(x)\right]
  - q(x)\int_0^x\rd z \frac{\rln^k(1-z)}{1-z}.
\end{equation}
Again using Eqs. (\ref{eq:steps}) and (\ref{eq:interp}) we have
\begin{eqnarray}
 P_kf_{k+}\otimes_xq & = &
           P_k\Biggl\{\sum_j\biggl[
     W_1\int_{x_j}^{x_{j+1}}\rd z\frac{\rln^{k}(1-z)}{z(1-z)}
   + W_2\int_{x_j}^{x_{j+1}}\rd z\frac{\rln^{k}(1-z)}{z^2(1-z)}
              \biggr] \nonumber \\
 & & \nonumber \\
 & & \hspace*{7mm} - q(x)\int_0^1\rd z\frac{\rln^{k}(1-z)}{1-z}\Biggr\}
 \label{eq:pi_int}.
\end{eqnarray}

Because each of the integrals appearing in (\ref{eq:pi_int})
diverges if the integration region includes $z=1$, the contribution from the
last step of integration must be evaluated with care.  For the case of a grid of
$N$ steps in $x$, the procedure is to first perform the sum on the
right hand side of (\ref{eq:pi_int}) up to $j=N-1$ and to perform the
integral of the last term up to $z=x_N$ only.  To obtain the contribution
from the $N^{th}$ $x$ step, $x_N<z<1$, the three terms on the right hand side of
(\ref{eq:pi_int}) are first combined.
The $\frac{1}{1-z}$ does not appear in the resulting expression, and so the
integral can be explicitly performed.  For $k=0$ and $k=1$
the contributions from the $N^{th}$ $x$ step are
\begin{equation}
  P_0\left[f_{0+}\otimes_xq\right]_N  = 
 P_0\left\{-q(x)\left[1+\rln(x_N)\right]+q\left(\frac{x}{x_N}\right)\right\}
   \nonumber \\
\end{equation}
and
\begin{eqnarray}
  P_1\left[f_{1+}\otimes_xq\right]_N & = &
 P_1\biggl\{\frac{x_N}{1-x_N}\left[-q(x)+q\left(\frac{x}{x_N}\right)\right]
            \left[\frac{1-x_N}{x_N}\rln(1-x_N)+\rln(x_N)\right] \nonumber \\
  & & \nonumber \\
 & & \hspace*{7mm}+ q(x)\left[-\rLi_2(1) + \rLi_2(x_N)\right]\biggr\}.
 \label{eq:I_int}
\end{eqnarray}

Because all of the parton distributions vanish at $x=1$, their approximation
with a linear interpolation,
in addition to improving the accuracy, is responsible
for the cancellation of the $\frac{1}{1-z}$ divergence.
Assembling the pieces from Eqs. (\ref{eq:p0_int}), (\ref{eq:pi_int}) and
(\ref{eq:I_int}), and including the contribution from the delta function term in
Eq. (\ref{eq:qcd_func}), gives the semianalytic expression for the convolution
integrals.\footnote[2]{Due to a typographical error in the presentation of this result
in reference \cite{thesis}, the full expression is repeated in appendix
\ref{ap:convolution}.}

In \cite{thesis}, a detailed comparison of the performance of the brute force
and semianalytic methods was performed for $g_1$ in the range
$0.0045 < x < 1.0$ and $1$ GeV$^2 < \qq < 60$ GeV$^2$.  As the number of $x$
steps was increased, both methods
converged on the same result.  To obtain the same accuracy within a few
tenths of a percent, which is sufficient for the $g_1$ data, 30 ${\rm ln}(\qq)$
steps and
40 (1280) ${\rm ln}(x)$ steps were required for the semianalytic (brute force)
methods.  The semianalytic evolution was about 350 times more efficient than
the brute force method, requiring approximately 4 seconds to
complete.\footnote[3]{Both methods were run on a SUN SPARC 10 workstation when
making this comparison.}
The following section on performance will focus on the convergence of the
semianalytic method and on the behavior of the structure
functions and evolved parton distributions.

\section{Performance}

\begin{figure}
\begin{center}
\mbox{\epsfxsize=\textwidth\epsffile{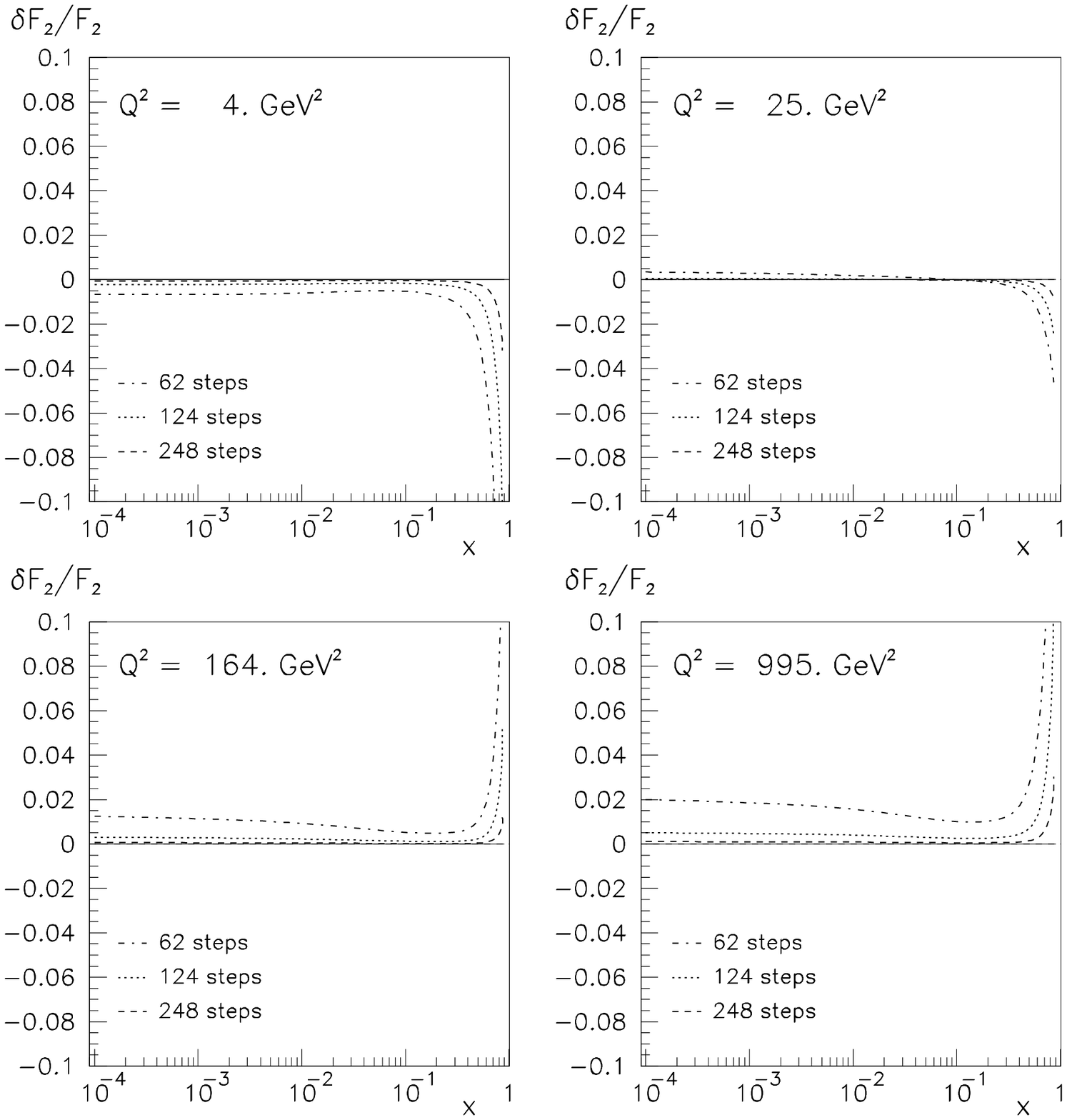}}
\end{center}
\caption[.]{\label{fig:f2rat}
 Results for $F_2(x)$ from the evolution procedure described in the text at
 four $\qq$ values.  The deviation of various results from a reference result,
 normalized to the reference result, is plotted.  The reference
 result was calculated with 496 ln($x$) steps and 200 ln($\qq$) steps.  The
 number of $x$
 steps used to obtain the other results is indicated on the figure.
 The MRS(A)~\cite{mrs_94} set of parton distributions at 4 GeV$^2$ was used
 as input.}
\end{figure}

\begin{figure}
\begin{center}
\mbox{\epsfxsize=\textwidth\epsffile{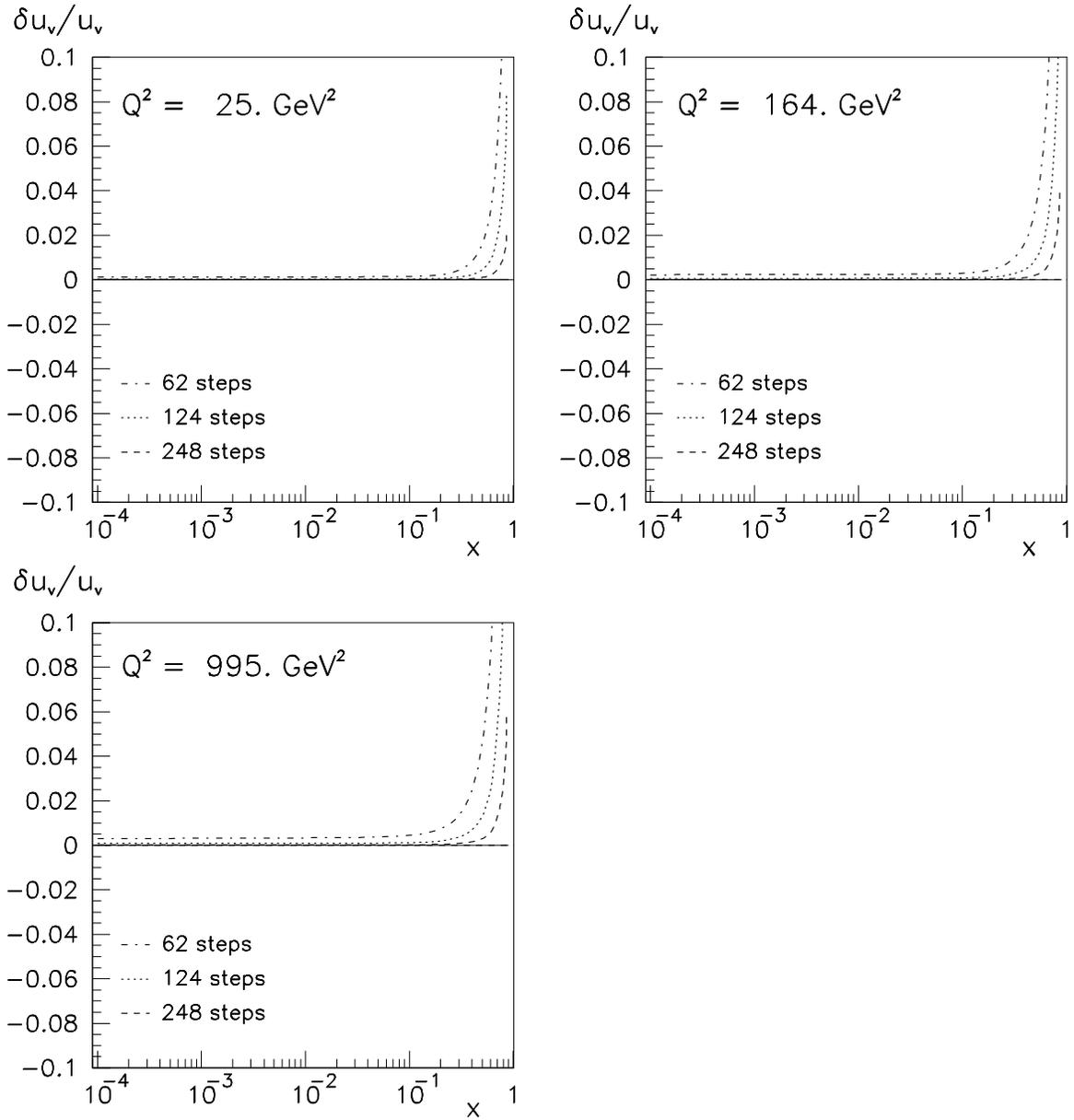}}
\end{center}
\caption{
 \label{fig:uvrat}
 \strut
 As in Fig. 1, but for the distribution $u_v(x)$.}
\end{figure}

\begin{figure}
\begin{center}
\mbox{\epsfxsize=\textwidth\epsffile{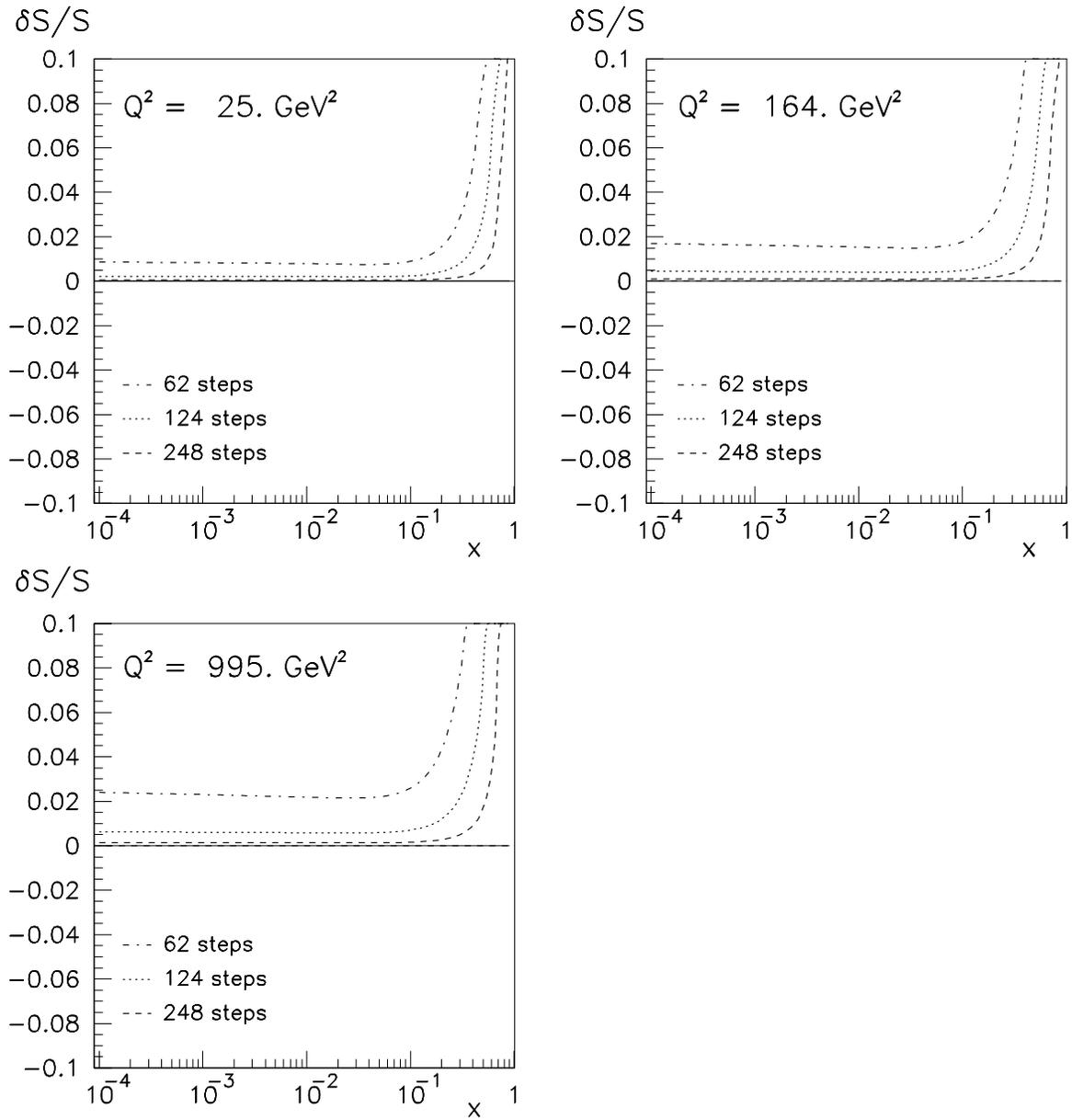}}
\end{center}
\caption{
 \label{fig:serat}
 \strut
 As in Fig. 1, but for $S(x)$, the total sea quark distribution.}
\end{figure}

\begin{figure}
\begin{center}
\mbox{\epsfxsize=\textwidth\epsffile{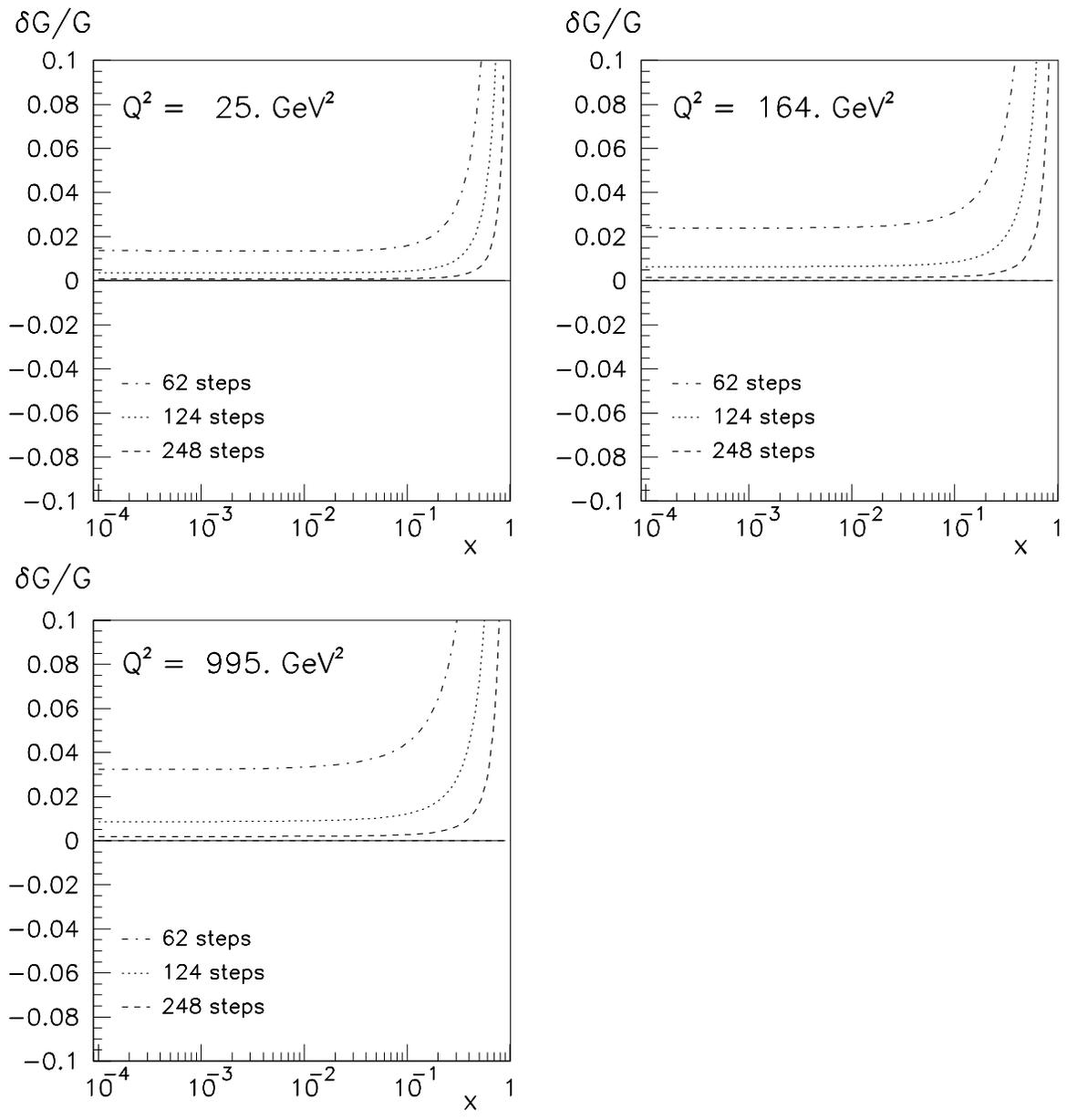}}
\end{center}
\caption{
 \label{fig:glrat}
 \strut
 As in Fig. 1, but for $G(x)$, the gluon distribution.}
\end{figure}

For the case of $F_2$ the MRS(A) set of distributions \cite{mrs_94},
defined in the $\overline{MS}$ renormalization scheme, was used as input
to the evolution.  The starting scale was $4.0$ GeV$^2$
and the distributions were evaluated down to $x=10^{-4}$ in the range
$4.0$ GeV$^2 < \qq < 10^4$ GeV$^2$.  The evolution was performed in
200 logarithmically distributed $\qq$ steps.  Several different numbers of
$x$ steps, also distributed logarithmically, were tried.  A reference
set of distributions was first calculated using 496 $x$ steps.  Distributions
calculated with 248, 124, and 62 $x$ steps were then compared with
these.  The results are shown in Figs. \ref{fig:f2rat}-\ref{fig:glrat} for
$F_2(x)$, $u_v(x)$, $S(x)$ and $G(x)$ where $u_v(x)$ is the valence up quark
distribution and $S(x)$ is the {\it total} sea quark
distribution.

In these figures, the difference between the reference set and the other sets
has been normalized to the reference set and plotted as a function of $x$ at
various $\qq$ values.  Note that all the quantities converge rapidly in the
region from $x = 10^{-4}$ up to some high value, which depends on the quantity
in question.  Beyond this value, there
is a strong dependence of the result on the number of $x$ steps.  Because the
spacing in $x$ is logarithmic, this is not surprising.  The situation could
presumably be improved by increasing the density of points in the high $x$
region.  For example, there are 9 points with $x>0.5$ for the case of 124
steps.  Comparing the 124 step and 248 step results for $F_2$, we see that
doubling this number to 18 would produce the same accuracy at $x=0.7$ as was
previously obtained at $x=0.5$.  This is a modest increase in the total number
of steps and increases the running time from just under to just over two
minutes.  In \cite{kumano_91} a linear step size was used at high $x$.

\begin{figure}
\begin{center}
\mbox{\epsfxsize=\textwidth\epsffile{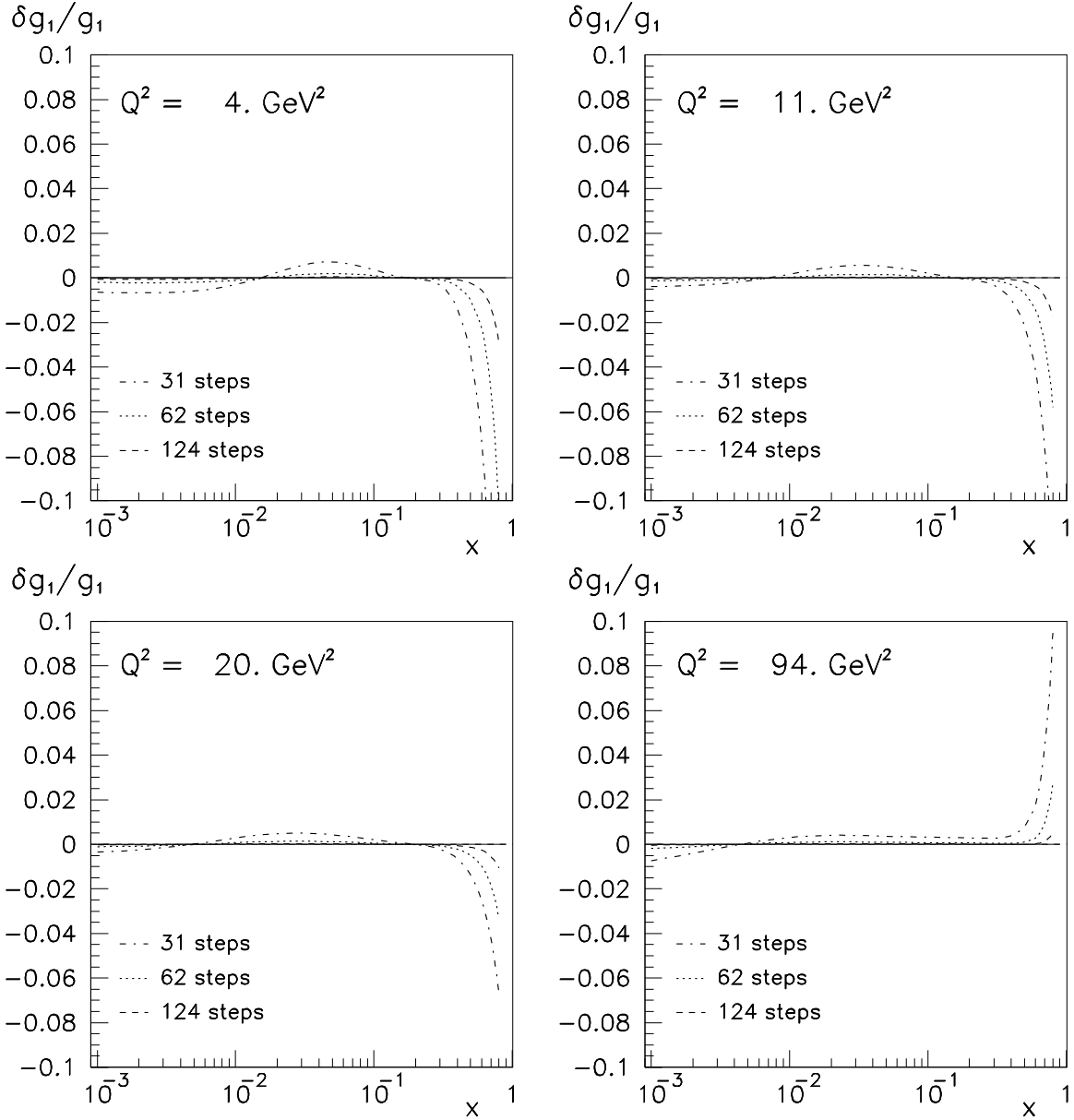}}
\end{center}
\caption[.]{\label{fig:g1rat}
 As in Fig. 1, but for the structure function $g_1(x)$.  In this case the
 reference result was calculated with 248 ln($x$) steps and 50 ln($\qq$)
 steps using the parton distributions of set A from~\cite{sg_96} at 4 GeV$^2$ as
 input.}
\end{figure}

Similar statements can be made for the parton distributions, though
for the gluon and sea quark distributions
convergence on an accurate result is not as rapid as
for $F_2$ and $u_v$.  However, if one is performing a fit of parton
distributions, it is only the accuracy with which $F_2$ is extracted which
will influence the fit parameters.  The convergence of the parton
distributions is not directly relevant to selecting a step size in $x$ and
therefore does not influence the time required to perform the fit.
If one only wants to evolve a fixed set of distributions to a high accuracy,
without the iterations necessary to perform a fit, a large number of
$x$ steps can be used while still keeping a reasonable running time.
In Fig. \ref{fig:g1rat} a similar accuracy study is presented for $g_1$.

\begin{figure}
\begin{center}
\mbox{\epsfxsize=\textwidth\epsffile{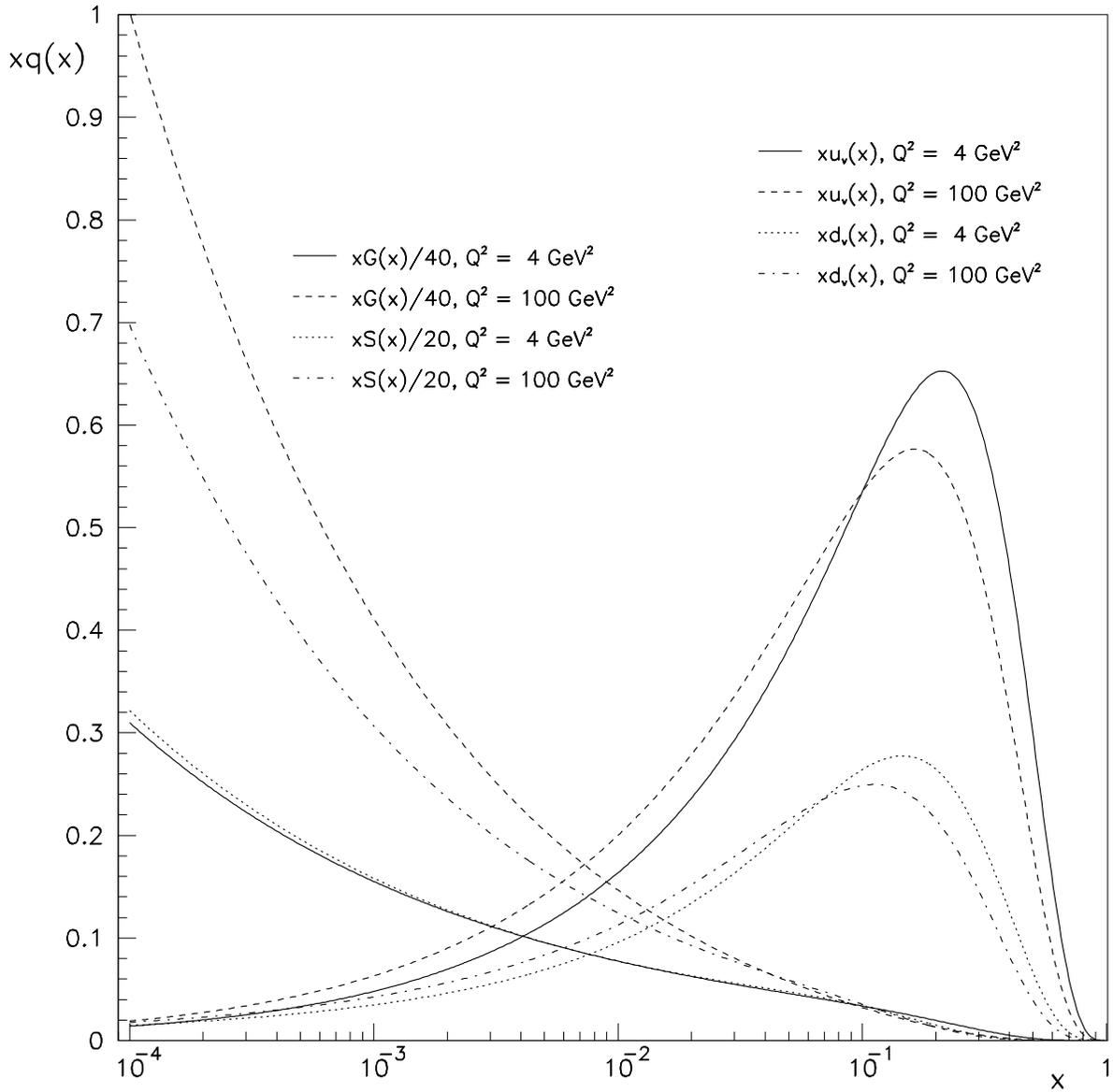}}
\end{center}
\caption{
 \label{fig:partons}
 \strut
 Parton distributions of the proton at the starting scale of 4 GeV$^2$
 and evolved to 100 GeV$^2$.}
\end{figure}

\begin{figure}
\begin{center}
\mbox{\epsfxsize=\textwidth\epsffile{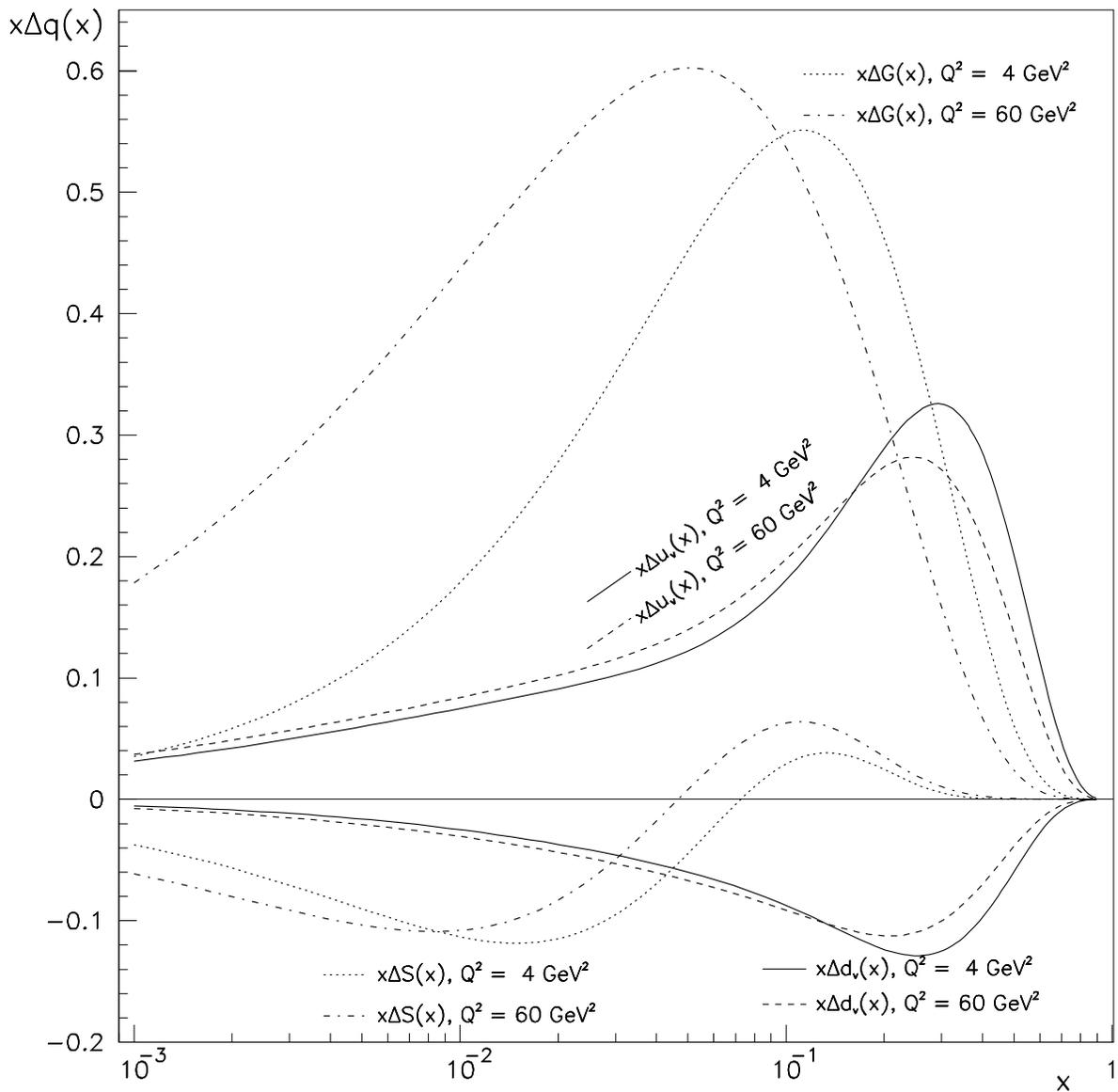}}
\end{center}
\caption[.]{\label{fig:dpartons}
 The polarization of the parton distributions of the proton at the starting
 scale of 4 GeV$^2$ and evolved to 60 GeV$^2$.}
\end{figure}

\begin{figure}
\begin{center}
\mbox{\epsfxsize=\textwidth\epsffile{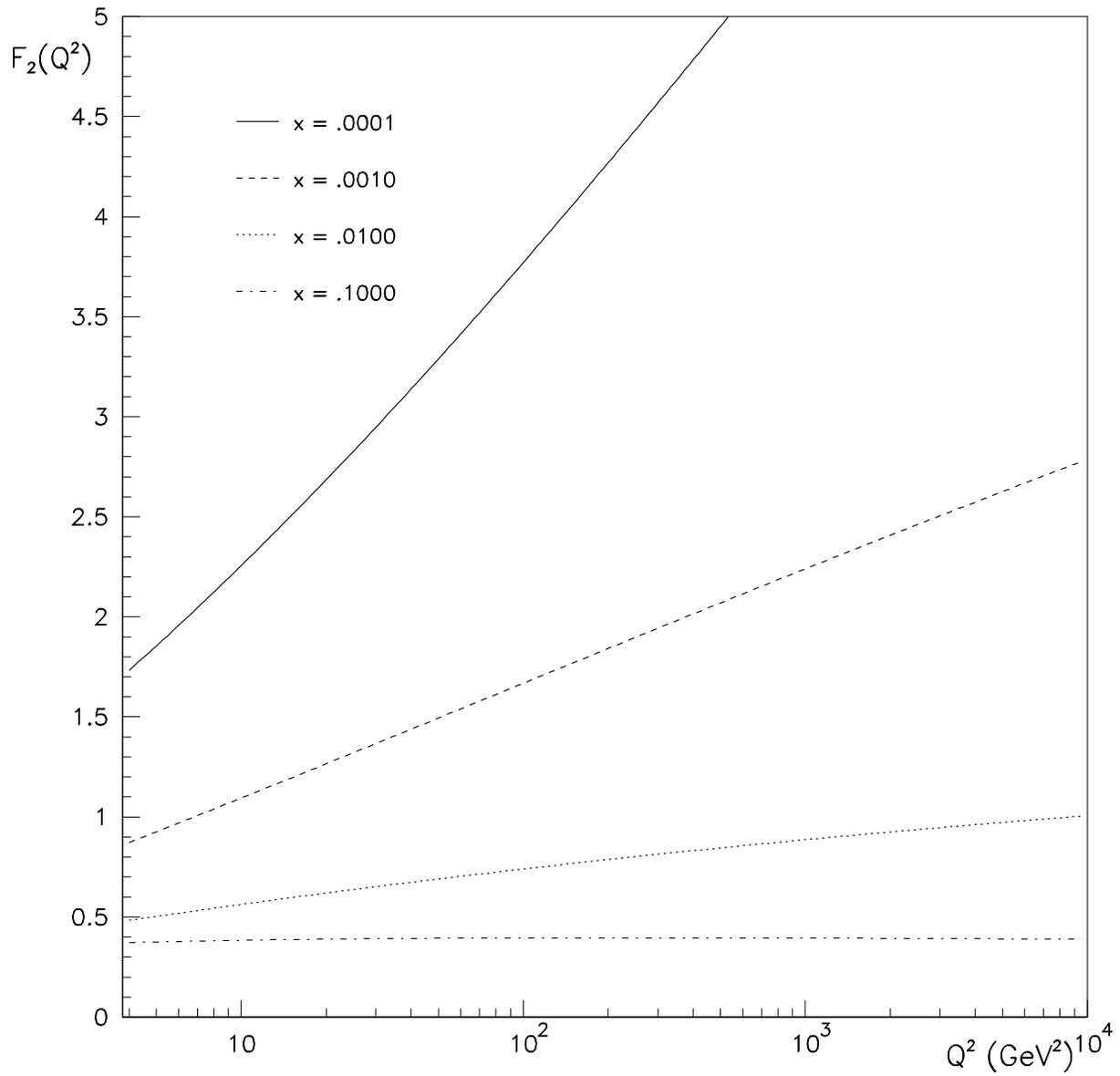}}
\end{center}
\caption{
 \label{fig:f2qq}
 \strut
 $F_2(\qq)$ at four $x$ values.}
\end{figure}

\begin{figure}
\begin{center}
\mbox{\epsfxsize=\textwidth\epsffile{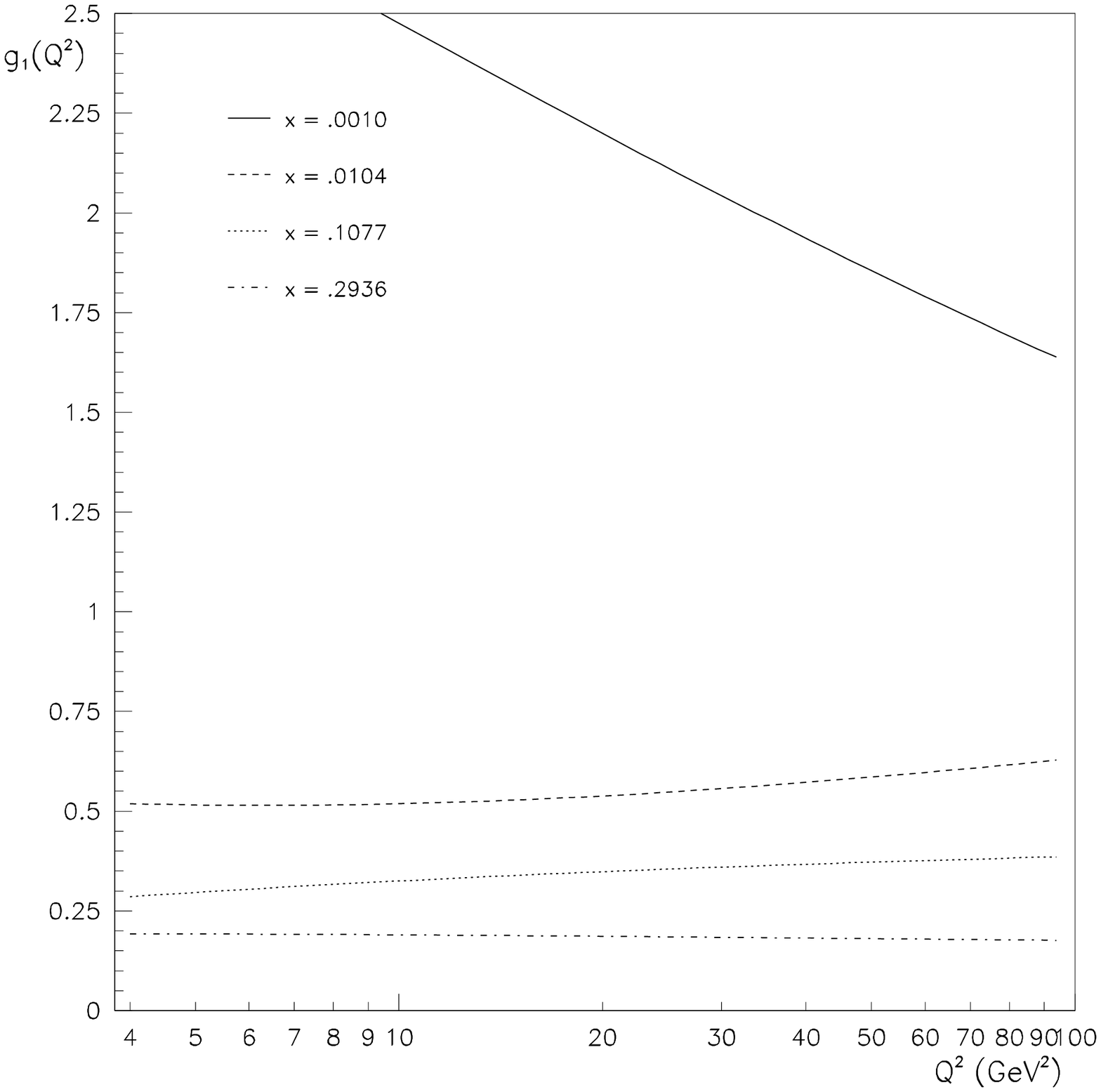}}
\end{center}
\caption{
 \label{fig:g1qq}
 \strut
 $g_1(\qq)$ at four $x$ values.}
\end{figure}

\begin{figure}
\begin{center}
\mbox{\epsfxsize=\textwidth\epsffile{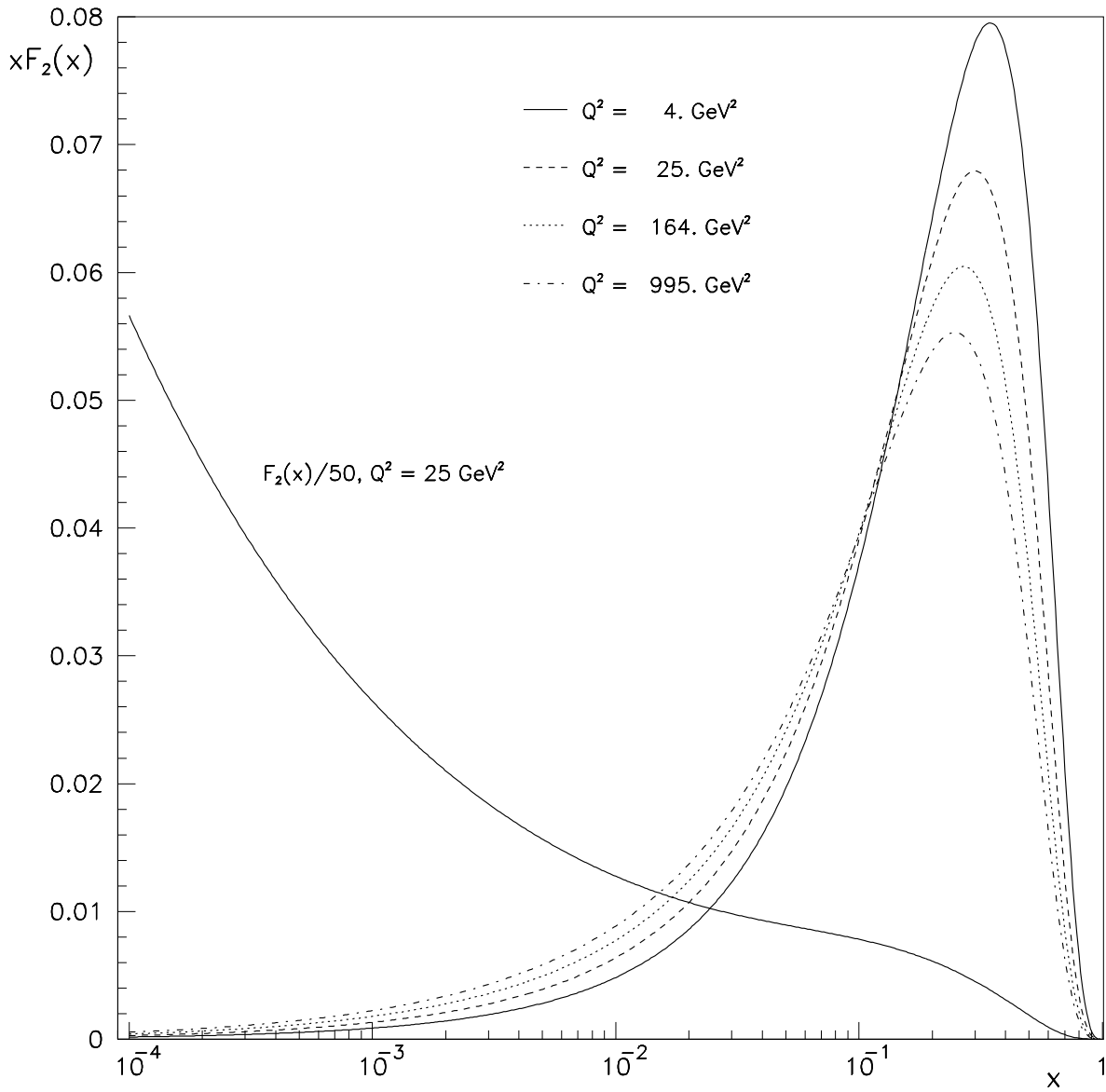}}
\end{center}
\caption{
 \label{fig:f2x}
 \strut
 $xF_2(x)$ at four $\qq$ values.  Also plotted is $F_2(x,25$ GeV$^2$).}
\end{figure}

\begin{figure}
\begin{center}
\mbox{\epsfxsize=\textwidth\epsffile{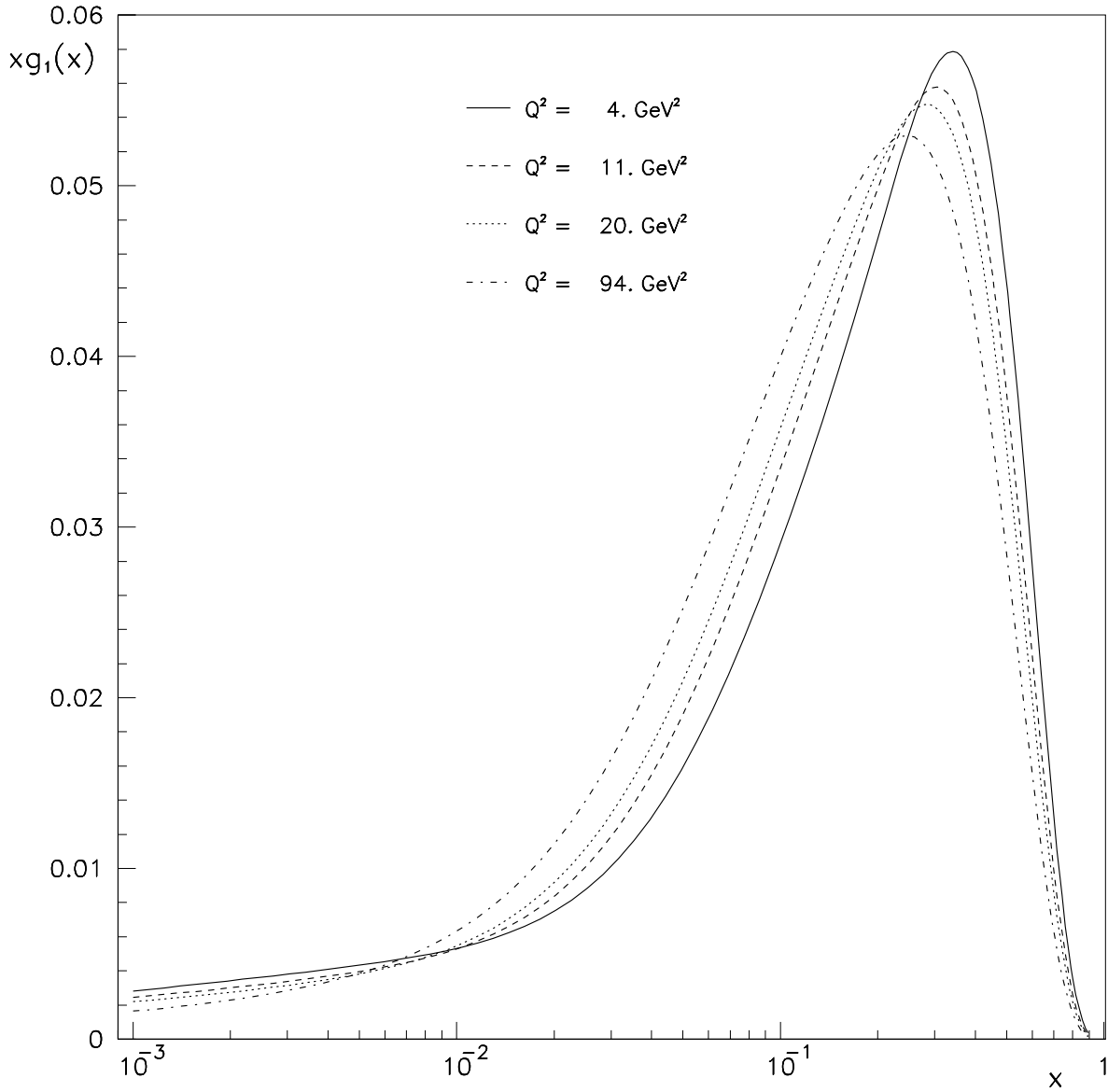}}
\end{center}
\caption{
 \label{fig:g1x}
 \strut
 $xg_1(x)$ at four $\qq$ values.}
\end{figure}

Figs. \ref{fig:partons} and \ref{fig:dpartons} show both the polarized and
unpolarized parton distributions at two values of $\qq$.  For the polarized
distributions, set A from \cite{sg_96} were used as input.  As in the
unpolarized case, the starting distributions were defined at $4$ GeV$^2$
in the $\overline{MS}$ scheme.  The polarized distributions were evaluated
down to $x=10^{-3}$ in the range $4$ GeV$^2 < \qq < 100$ GeV$^2$
with 62 ${\rm ln}(x)$ steps and 50 ${\rm ln}(\qq)$ steps.  The unpolarized
distributions which are displayed in Fig. \ref{fig:partons}
were evaluated with 496 ln($x$) steps and 200 ln($\qq$) steps.
The structure functions subsequently derived from these parton
distributions are shown in Figs. \ref{fig:f2qq}-\ref{fig:g1x}.  In Figs.
\ref{fig:f2qq} and \ref{fig:g1qq} one can directly see the logarithmic $\qq$
behavior of the structure functions at various values of $x$.  A comparison
shows that the sizes of the slopes normalized to the structure function
value are similar for the two structure functions.  At the lowest
values of $x$ the behavior of the two structure functions is quite different;
both have slopes which are similar in magnitude but opposite in sign.  This is
consistent with expectations, the density of partons in the low $x$ region
being dominated by the sea quark and gluon distributions.
As the resolution of the probe increases,
more partons are seen, but their helicity is not always remembered by the
processes which generate them.  Because the $\qq$ behavior of $F_2$ and $F_1$
are similar, these plots are also a dramatic illustration of the danger in the
low $x$ region of assuming $\frac{g_1}{F_1}$ is independent of $\qq$, as has
been done in evaluating the first moment of $g_1(x,\qq)$,
$\Gamma_1(\qq)$ \cite{smc,slac}.  The differing low $x$ behavior of these
structure functions is evident from another point of view in
the plots of $xF_2(x)$ and $xg_1(x)$ in Figs. \ref{fig:f2x} and \ref{fig:g1x}.
Here we can also see that the
$\qq$ slope of $F_2$ turns over around $x=0.12$, which is just above the
highest $x$ value plotted in Fig. \ref{fig:f2qq}.  The $\qq$ slope of $g_1$
turns over twice, behaving as $F_2$ in the valence region but becoming negative
where the sea dominates.
Finally, in Fig. \ref{fig:f2x}
$F_2(x,25 $GeV$^2)$ (scaled by a factor of 50) is
also plotted in order to facilitate a comparison with the same quantity plotted
in reference \cite{mrs_hepph}.

\section{Conclusions}

A rapid method for the evolution of parton distributions in $x$ space has been
presented.  As a numerical technique, it places no restrictions on the
form of the distributions to be evolved.  This freedom, as well as its ability
to accommodate the evolution equations modified for recombination effects, may
prove useful as the kinematic range of the data grows and its statistical
accuracy improves.  Pending a careful numerical analysis of the results,
one can see now that the evolved distributions and the structure functions
presented here compare favorably with similar figures presented in references
\cite{sg_96} and
\cite{mrs_hepph}.\footnote[4]{The code may be obtained from the author via email
at fasching@wisconsin.cern.ch.}
\newline
\vskip 1mm
\hskip -6mm\begin{underline}{\bf Acknowledgments}
\end{underline}
\newline
\vskip .1mm
\hskip -6mm Many thanks to Mayda Velasco and Stephen Trentalange for
proofreading this text and for many useful comments.   Kind thanks also to
W.J. Stirling and T. Gehrmann for providing the FORTRAN code for the parton
distributions used here.

\appendix

\section{Form of the Convolution Integral in the Semianalytic Method}
\label{ap:convolution}

Eq. (\ref{eq:conv_full}) gives the full expression of the semianalytic
convolution of a splitting function or coefficient function of the form
\begin{equation}
 P(x) = P_a(x) + \sum_{k=0}^1 P_k\cdot\left(\frac{{\rm ln}^k(1-x)}{1-x}\right)_+
        + P_\delta\delta(1-x),
\end{equation}
with a parton distribution approximated on an $x$ grid by
\begin{equation}
 q\left(\frac{x_i}{z}\right)_j =
    \frac{q\left(\frac{x_i}{x_{j+1}}\right) - q\left(\frac{x_i}{x_j}\right)}
         {\left(\frac{x_i}{x_{j+1}} - \frac{x_i}{x_j}\right)} \cdot
    \left(\frac{x_i}{z} - \frac{x_i}{x_j}\right)
 + q\left(\frac{x_i}{x_j}\right),
 \label{eq:ap_part}
\end{equation}
where $j$ is the $x$ grid index,
$x_j$ and $x_{j+1}$ are the upper and lower boundaries of step $j$ and $x_i$ is
the lower limit of the integration region.  If the
integration region has been divided into $N$ steps in $x$, with $x_{N+1} = 1$,
we have

\begin{eqnarray}
   P\otimes_{x_i}q & = &  P_\delta q(x_i) \nonumber \\
 & & + \sum_{j=i}^N\int_{x_j}^{x_{j+1}}
		 \biggl[ W_1\frac{P_a(z)}{z}
                + W_2\frac{P_a(z)}{z^2}\biggr]{\rm d}z
			     \nonumber \\
 & & + \sum_{k=0}^{1}P_k\sum_{j=i}^{N-1}\int_{x_j}^{x_{j+1}}
		  \biggl[W_1\frac{{\rm ln}^k(1-z)}{z(1-z)}
		+ W_2\frac{{\rm ln}^k(1-z)}{z^2(1-z)}
                - q(x_i)\frac{{\rm ln}^k(1-z)}{1-z}\biggr]{\rm d}z
			     \nonumber \\
 & & + P_0 \left\{- q(x_i)\left[1+{\rm ln}(x_N)\right]
        + q\left(\frac{x_i}{x_N}\right) + q(x_i){\rm ln}(1-x_i)\right\}
			\nonumber \\
 & & + P_1 \biggl\{\frac{x_N}{1-x_N}
                    \left(-q(x_i)+q\left(\frac{x_i}{x_N}\right)\right)
            \left(\frac{1-x_N}{x_N}{\rm ln}(1-x_N)+{\rm ln}(x_N)\right)
			      \nonumber \\
 & & \hspace*{10mm} + q(x_i)\left(-{\rm Li}_2(1) + {\rm Li}_2(x_N)\right)
        + \frac{1}{2}q(x_i){\rm ln}^2(1-x_i)\biggr\},
 \label{eq:conv_full}
\end{eqnarray}
where $W_0$ and $W_1$ are defined in Eqs. (\ref{eq:w1}) and (\ref{eq:w2}).
Evaluating a structure function at NLO at a single $\qq$ value involves
evaluating (\ref{eq:conv_full}) for each of the $N$ values of $x_i$ for
each of the involved splitting and coefficient functions.

\end{document}